\documentclass{tPHL2e}

\usepackage{epstopdf}
\usepackage{subfigure}

\theoremstyle{plain}

\theoremstyle{definition}

\theoremstyle{remark}

\begin{document}

\title{\textit{Evidence for field induced proximity type behavior in $CoFe_2O_4$ based ferromagnetic nanofluid}}

\author{
\name{S. Sergeenkov\textsuperscript{a}$^{\ast}$\thanks{$^\ast$Corresponding author. Email: sergei@df.ufscar.br}, 
C. Stan\textsuperscript{b}, C.P. Cristescu\textsuperscript{b}, M. Balasoiu\textsuperscript{c,d}, N.S. Perov\textsuperscript{e} and  C. Furtado\textsuperscript{a}}
\affil{\textsuperscript{a}Department of Physics, Universidade Federal da Para\'{\i}ba, Jo\~{a}o Pessoa, PB, Brazil; \textsuperscript{b}Department of Physics, Faculty of Applied Physics, Politehnica University of Bucharest, Romania; \textsuperscript{c}Joint Institute for Nuclear Research, Dubna, Russia; \textsuperscript{d}Horia Hulubei National Institute of Physics and Engineering, Romania; \textsuperscript{e}Faculty of Physics, Lomonosov Moscow State University, Moscow, Russia}}


\maketitle

\begin{abstract}
We report some unusual magnetic properties observed in $CoFe_2O_4$ based ferrofluid (with an average particle size of $D=6nm$). More precisely, in addition to the low-field ferromagnetic (FM) phase transition with an  intrinsic Curie temperature $T_{Cb}=350K$, a second phase transition with an extrinsic Curie temperature $T_{Cw}=266K$ emerges at higher (saturating) magnetic field. The transitions meet at the crossover point $T_{cr}=210K$. The origin of the second transition is attributed to magnetic field induced proximity type interaction between FM particles through non-FM layers.
\end{abstract}


\section{Introduction}

Due to a large scale applicability, the nanostructured materials continue to attract considerable attention. Of special importance are the stable colloidal dispersions of ferromagnetic (FM) nanoparticles [1-5] which  are also of interest both for experimental and theoretical studies related to the influence of different size effects on their structural and magnetic properties [6-10]. In particular, it was found [9,10] that the so-called finite temperature size effects can significantly reduce the amount of the ordered phase and the corresponding value of the Curie temperature. 
A high coercivity at room temperature,  moderate saturation magnetization and good catalytic properties made cobalt ferrite ($CoFe_2O_4$) nanoparticles not only an ideal material for high density storage devices, photomagnetic applications, and molecular agents in magnetic resonance imaging but also proved them one of the most popular ingredients of ferrofluids [11-14]. It is now well established that the overall behavior of ferromagnetic fluids is determined by a competition between their intrinsic  (intraparticle)  and extrinsic (interparticle) properties. However, a separate study of these two contributions still remains a challenging problem. 
On the other hand, it should be pointed out that despite the fact that ferrofluids are based on such a strong FM material as $CoFe_2O_4$ (with the Curie temperature as much as $T_{C}=793K$), their magnetic response is mistakenly treated within the Langevin model of paramagnetism (PM) which considers the magnetic fluid as a system composed by non-interacting spheres with a permanent dipolar magnetic moment. So, when an external magnetic field is applied, the magnetic dipole moment of each nanoparticle experiences a torque which makes it rotate around the direction of the field. Therefore, the macroscopic magnetization within the Langevin scenario is the result of the combined action of magnetic moments orientation induced by the external field and the perturbation produced by the disordered Brownian motion (thermally activated diffusion) [6,11-13]. Needless to say that within this approach the most interesting feature of ferrofluids (related to the ordering of the FM particles below the corresponding Curie temparatures) is completely missing from the picture.  
In this regard, it is imperative to mention earlier theoretical attempts to model phase behavior of the ferrofluid based on the Curie-Weiss (rather than Langevin) concept by taking into account realistic interactions between FM particles within the fluid [15-17].  
To properly address this important issue, in this Letter we present our latest results on the magnetic field and temperature behavior of $CoFe_2O_4$ based ferrofluids. We found that in addition to the major low-field phase transition (with the Curie temperature $T_{Cb}=350K$), a second phase transition (with the Curie temperature $T_{Cw}=266K$) is developing at higher (saturating) magnetic field. The origin of the second (extrinsic) transition is attributed to manifestation of strong field induced proximity type processes between the nearest particles in the ferrofluid. 
 
\begin{figure}
\centerline{\includegraphics[width=5cm]{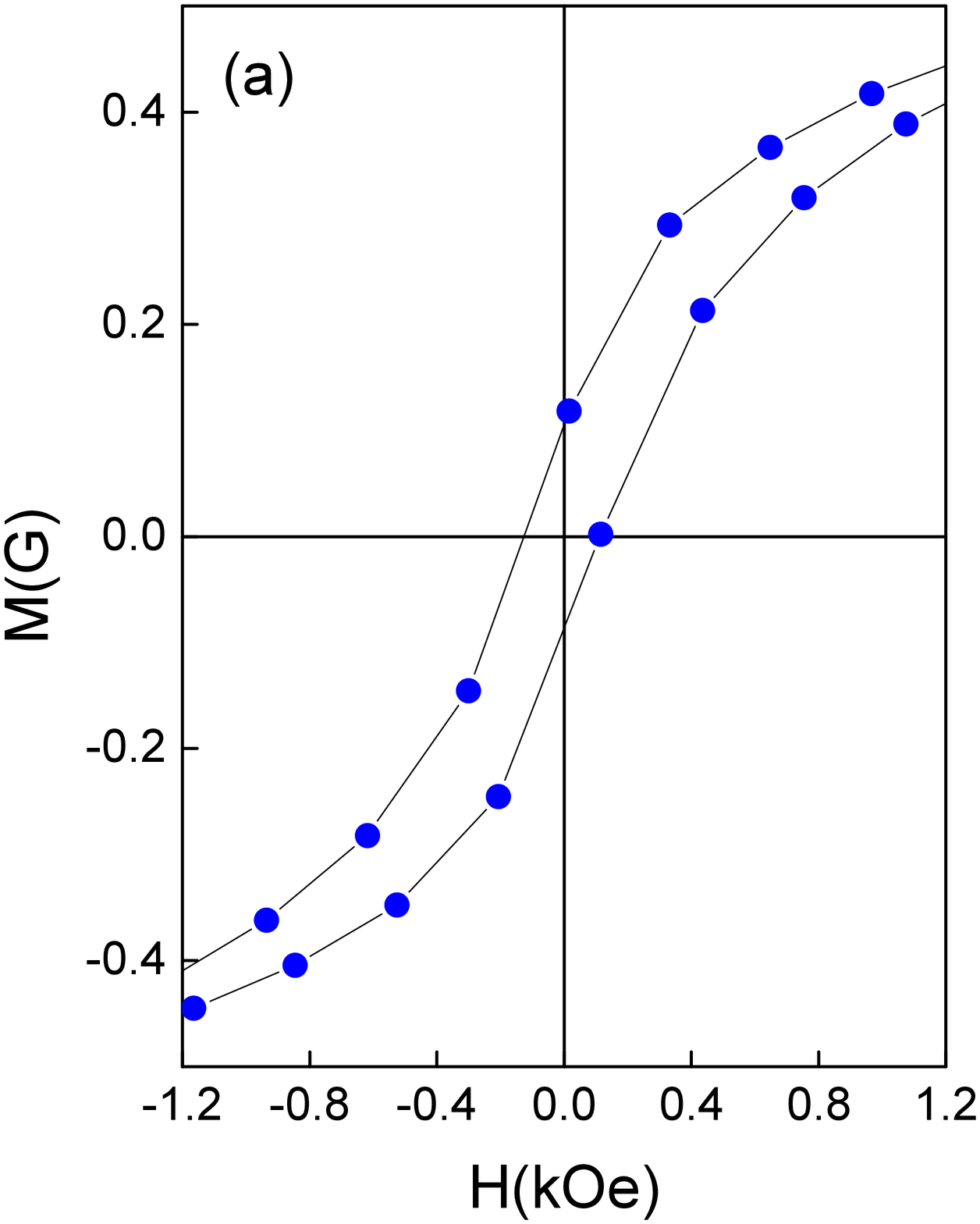}}\vspace{0.25cm}
\centerline{\includegraphics[width=5cm]{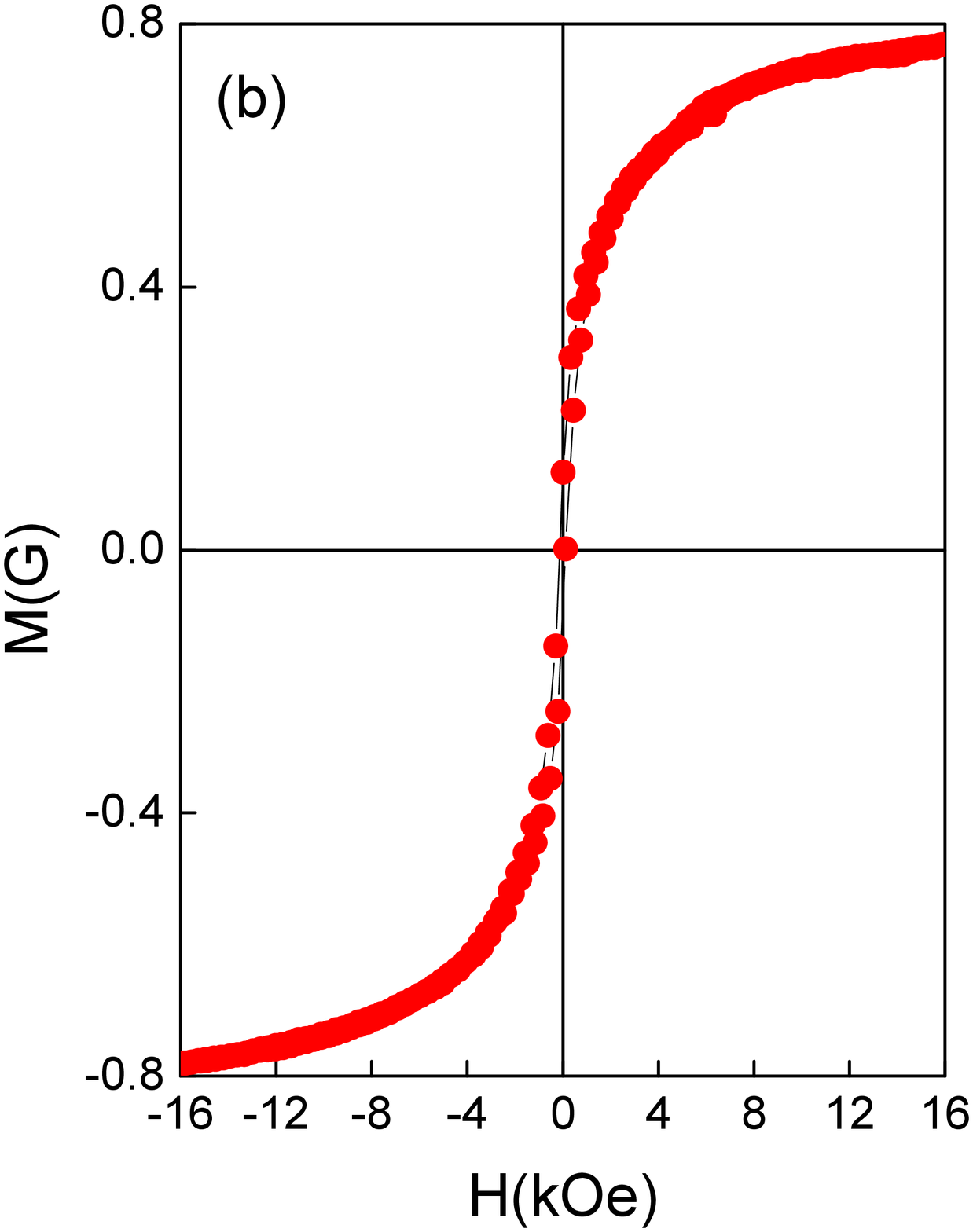}}
\vspace{0.25cm} \caption{$M-H$ curves taken at the highest temperature ($T=350K$) for low (a) and high (b) magnetic field regions.} 
\label{fig:fig1}
\end{figure}

\begin{figure}
\centerline{\includegraphics[width=5cm]{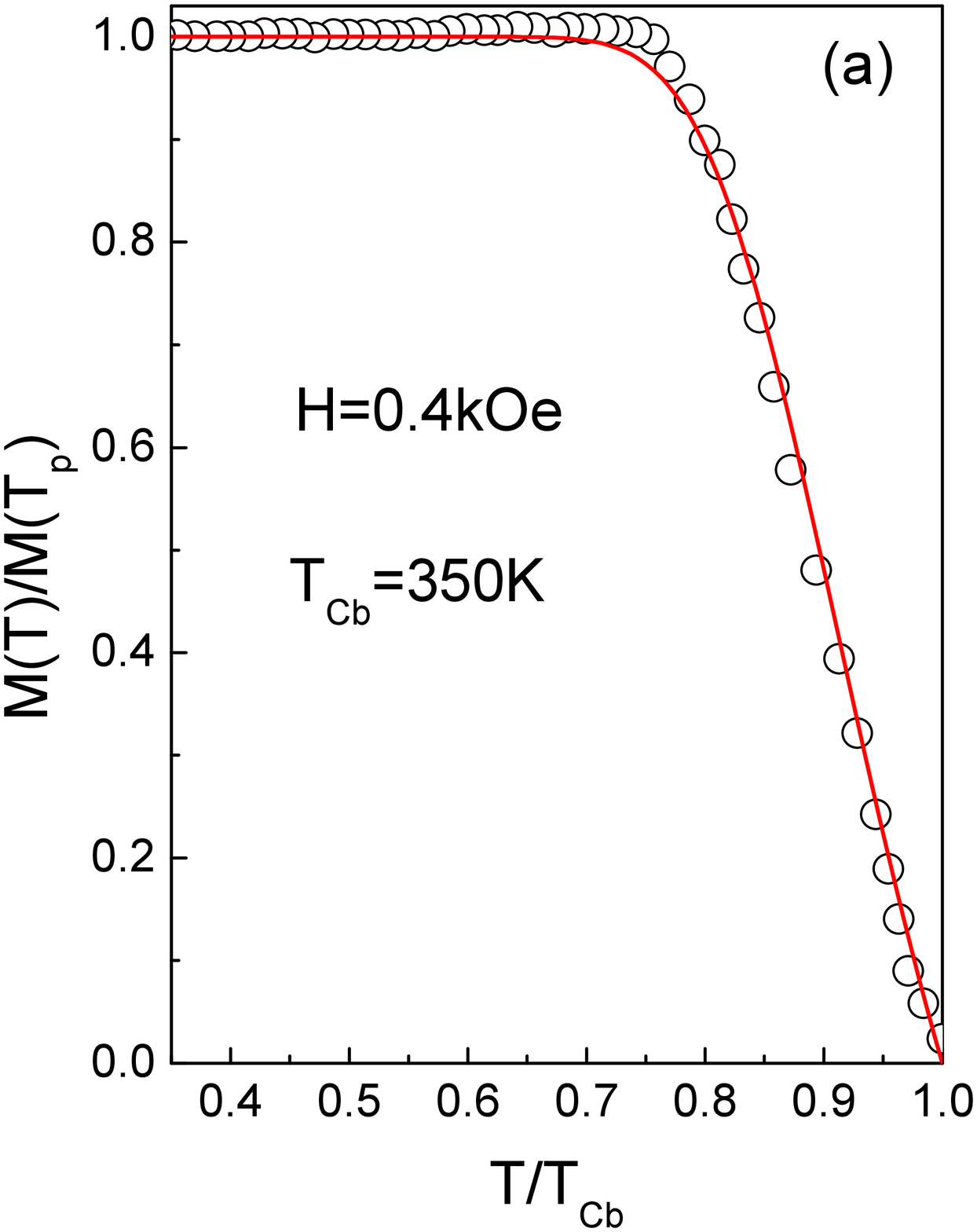}}\vspace{0.25cm}
\centerline{\includegraphics[width=5cm]{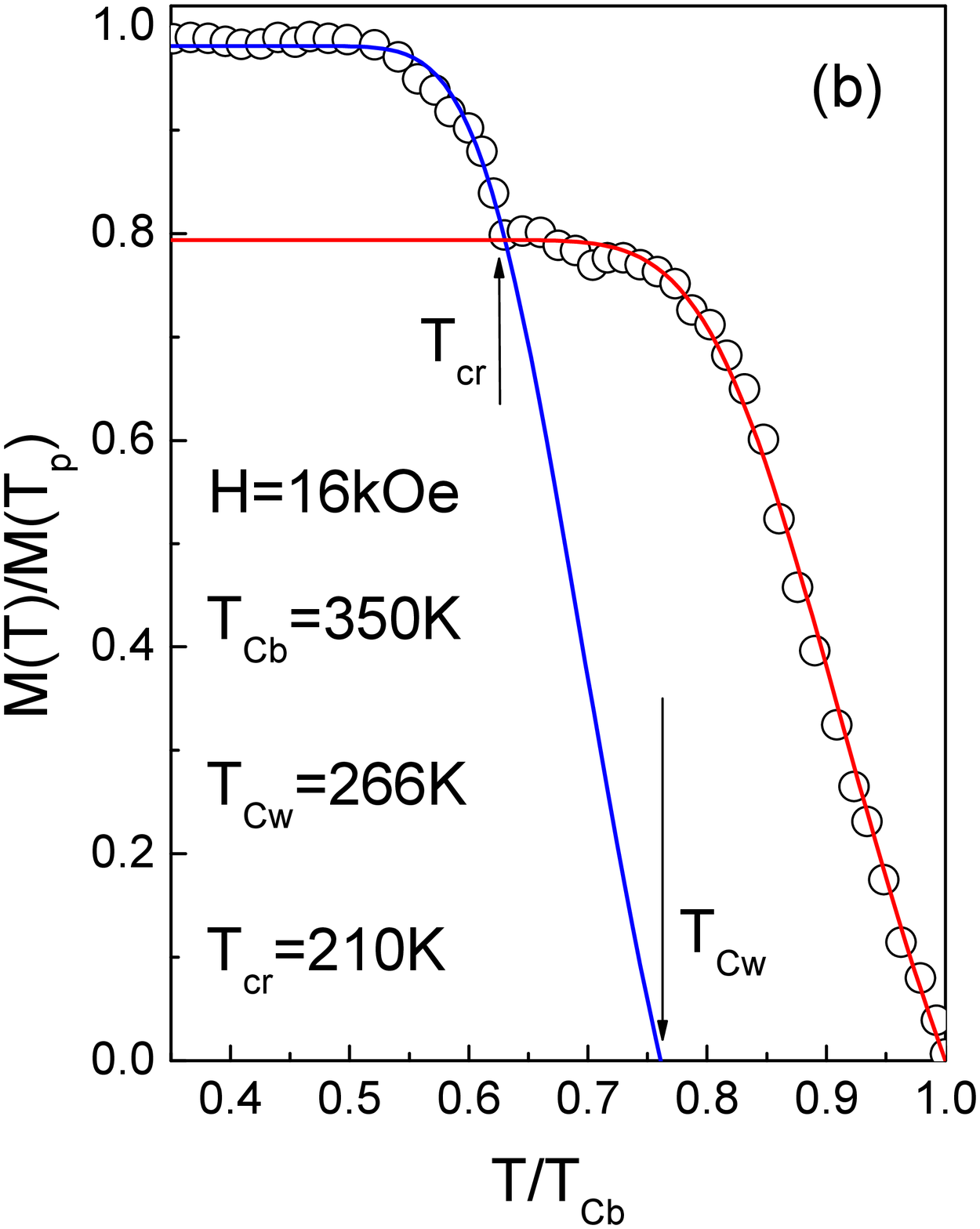}}\vspace{0.25cm}
\caption{ The  dependence of the normalized magnetization $M(T)/M(T_p)$ on the reduced temperature $T/T_{Cb}$ taken at low (a) and high (b) magnetic field. The solid lines are the best fits according to Eqs.(1)-(3).} 
\label{fig:fig2}
\end{figure}

\section{Samples characterization and magnetic measurements}

Samples of $CoFe_2O_4$ ferrofluid nanoparticles coated with a double layer of dodecylbenzenesulphonic acid (DBS) and dispersed in double distillated water were synthesized by chemical co-precipitation method presented in [11-13]. The structural characterization of our samples was performed by transmission electron microscopy (TEM) and small angle neutron scattering (SANS). The analysis of TEM images and SANS spectra confirmed [18] the presence of well-defined FM nanoparticles (with an average diameter of $D=6nm$) separated by non-FM layers (with an average thickness of $t=3.5nm$). Magnetic measurements were carried out on a LakeShore N7400 vibrating sample magnetometer in the temperature interval $110K \le T \le 350K$. Fig.\ref{fig:fig1} shows typical $M-H$ curves (taken at $T=350K$) for low  and high magnetic field regions. Notice that  low-field hysteresis is quite strong (with coercive magnetic field $H_C=0.2kOe$). 

\section{Results and discussion}

Fig.\ref{fig:fig2} presents the dependence of the normalized magnetization $M(T)/M(T_p)$ on the reduced temperature $T/T_{Cb}$ after subtraction of the paramagnetic contribution. Here, $M(T_p)=0.29G$ with $T_p=110K$.  More specifically, Fig.2(a) shows the temperature behavior of $M(T)$ under the low magnetic field $H=0.4kOe$.  Likewise, Fig.2(b) presents the temperature dependence of magnetization $M(T)$ taken under the higher magnetic field $H=16kOe$ with $T_p=110K$ and $M(T_p)=0.78G$. As we can see, in this case  $M(T)$ exhibits more interesting behavior. Namely, in addition to the major phase transition at $T_{Cb}=350K$, a second FM transition is emerging at $T_{Cw}=266K$. The inflection point at $T_{cr}=210K$ indicates a clear-cut crossover between the two transitions. Similar crossover takes place in room-temperature nanogranular ferromagnetic graphite [10] and in arrays of quantum dots [19].  
Turning to the analysis of the obtained results, let us begin with the low-field magnetization. 
We were able to successfully fit the data using the following explicit expression for the single particle (intrinsic) magnetization:
\begin{equation}
M_{b}(T)=M_{sb}\tanh\left[\left(\frac{T_{Cb}}{T}\right)^4-1\right]
\end{equation}
which presents approximate solution of the Curie-Weiss mean-field equation for FM magnetization valid for all temperatures [4,5,9,10]. The solid line in Fig.2(a) is the best fit according to Eq.(1) with $T_{Cb}=350K$ and $M_{sb}=0.25G$. Notice that, according to Fig.1(a), the value of $M_{sb}$ well correlates with $M(H=0.4kOe)$. 

As far as the high-field magnetization is concerned, it is quite natural to assume that in addition to intrinsic contribution $M(T)=M_{b}(T)$ (given by Eq.(1)) above the crossover point $T_{cr}$, the observed temperature behavior below $T_{cr}$ can be attributed to manifestation of the field induced second phase transition (with the extrinsic Curie temperature $T_{Cw}=266K$). More precisely, we were able to successfully fit all the high-field data using the following explicit expressions:
\begin{equation}
M_{b}(T)=M_{sb}\tanh\left[\left(\frac{T_{Cb}}{T}\right)^4-1\right], \qquad T>T_{cr}
\end{equation}
and
\begin{equation}
M_{w}(T)=M_{sw}\tanh\left[\left(\frac{T_{Cw}}{T}\right)^4-1\right], \qquad T<T_{cr}
\end{equation}
above and below the inflection point $T_{cr}$, respectively. 
The solid lines in Fig.2(b) present the best fits according to Eqs.(2) and (3) with $T_{Cb}=350K$, $T_{Cw}=266K$, $M_{sb}=0.62G$, and $M_{sw}=0.78G$. 

To understand why the low-field magnetization does not exhibit a crossover, let us elucidate the origin of the second phase transition. Recall that in granular materials the interparticle (intergranular) contribution $M_{w}(T)$ is governed by the tunneling modified exchange interaction between the nearest particles with probability $\propto e^{-d/2\xi}$ where $d=2t$ is the distance between neighboring particles ($t=3.5nm$ is the thickness of the non-FM coating layer around FM nanoparticle in our ferrofluid) and $\xi=h/\sqrt{2mU}$  is a characteristic length with $U$ being the potential barrier height,  $m$ an effective carrier mass and $h$ the Plank constant. 
As a result, the interparticle (extrinsic) Curie temperature $T_{Cw}$ can be related to the intraparticle (intrinsic) Curie temperature $T_{Cb}$ as follows [10,19]
\begin{equation}
T_{Cw}=e^{-d/2\xi}T_{Cb}
\end{equation}
Using the experimentally found values for $T_{Cb}$, $T_{Cw}$ and $d$, from Eq.(4) we obtain $\xi=10nm$ for an estimate of the characteristic (coherence) length. It is important to keep in mind that the potential barrier $U$ (and hence $\xi$) depends on applied magnetic field $H$. With a good enough accuracy, we can assume that $U(H)=U(0)-\mu H$ where $\mu$ is the total magnetic moment of the ferrofluid and a zero-field contribution $U(0)$ is dominated by the highest thermal energy which in our case is the bulk Curie temperature $T_{Cb}$, that is $U(0)\simeq k_BT_{Cb}$. At low magnetic fields (far from the saturation), $U(0)\gg \mu H$ and the potential barrier $U(H)$ is too high to provide an efficient tunneling between FM particles since in this case $\xi(H) \propto \frac{1}{\sqrt{U(H)}} \ll d$ leading to a strong reduction of the interparticle Curie temperature $T_{Cw}\ll T_{Cb}$ according to Eq.(4). On the other hand, at high enough magnetic fields (near the saturation), $U(0)\simeq \mu H$ which results in a significant lowering of the barrier $U(H)$ and provides a strong interaction between the particles forming the nanofluid. In this case, $\xi(H) \simeq  d$ leading to manifestation of the observed second phase transition at $T=T_{Cw}$. To better understand the difference between the low and high field situations, let us consider a particular example with $U(H_2)=0.04U(H_1)$ where $H_1=0.4kOe$ and $H_2=16kOe$. By definition, $\frac{\xi(H_1)}{\xi(H_2)}= \sqrt{\frac{U(H_2)}{U(H_1)}}$. So, in view of Eq.(4) and recalling that $d=7nm$ and $\xi(H_2)=10nm$, we obtain $\xi(H_1)\simeq 0.2\xi(H_2)\simeq 2nm$ and $T_{Cw}(H_1)\simeq 80K$ for low-field coherence length and the corresponding tunneling Curie temperature, respectively. This example demonstrates that at low fields (far below saturation) the interparticle contribution is shifted towards low temperatures making it more difficult to observe the proximity type effect in our samples.

\begin{figure}
\centerline{\includegraphics[width=5cm]{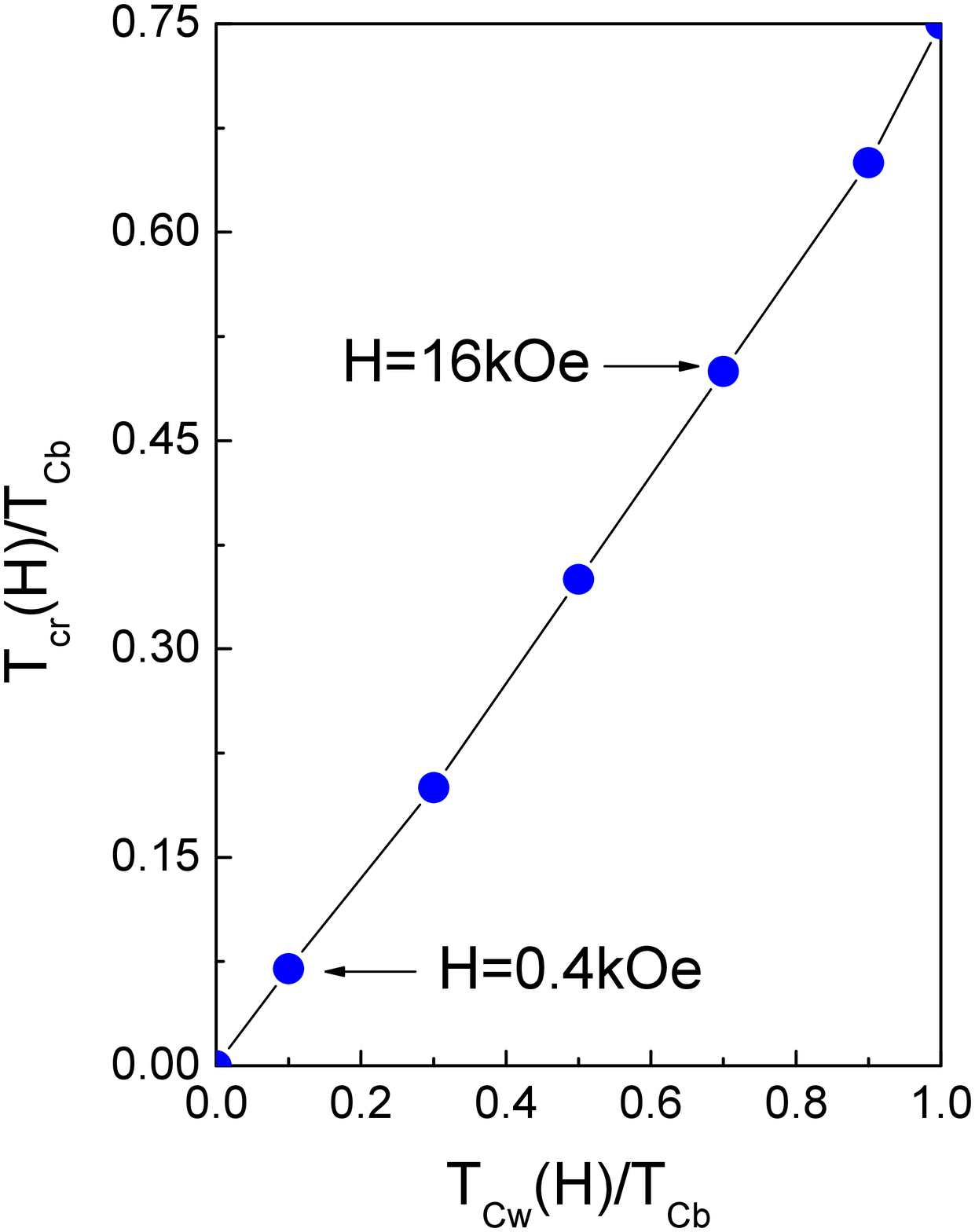}}\vspace{0.25cm}
\caption{ The dependence of the reduced crossover temperature $T_{cr}(H)/T_{Cb}$ on the ratio $T_{Cw}(H)/T_{Cb}$. The dots are the solutions of the equation $M_b(T_{cr})=M_w(T_{cr})$.} 
\label{fig:fig3}
\end{figure}

Let us discuss now the origin of the crossover temperature $T_{cr}$. Notice that unlike the bulk temperature $T_{Cb}$, its weak-links-mediated counterpart $T_{Cw}$ is strongly influenced by magnetic field. This becomes evident when we rewrite Eq.(4) as follows, $T_{Cw}(H)=e^{-d/2\xi(H)}T_{Cb}$.  According to Eqs.(2) and (3), $T_{cr}(H)$ is given by the solution of the equation $M_b(T_{cr})=M_w(T_{cr})$ for a fixed value of $T_{Cw}(H)$.  The result is presented in Fig.\ref{fig:fig3} which depicts the dependence of the reduced crossover temperature $T_{cr}(H)/T_{Cb}$ on the ratio $T_{Cw}(H)/T_{Cb}$.  As we can see, the value of the crossover temperature directly follows the value of the extrinsic Curie temperature, that is $T_{cr}(H) \propto T_{Cw}(H)$.  Two relevant reference points, corresponding to $H=0.4kOe$ and $H=16kOe$, are marked with arrows.

And finally, an important comment is in order regarding the absolute value of the single particle Curie temperature $T_{Cb}=350K$. Recall [20] that the Curie temperature of the bulk $CoFe_2O_4$ crystal can reach as much as $T_{C}=793K$. This seeming discrepancy has to do with a pronounced size effect in nanoparticles. To quantify this statement, we can relate these two temperatures using the concept of the thermal de Broglie wavelength $\Lambda=h/\sqrt{mk_BT}$ which is responsible for finite temperature size effects in nanogranular materials [10]. More explicitly, one of such relationships between $T_{C}$ and $T_{Cb}$ reads:
\begin{equation}
T_{Cb}=\left(\frac{D}{D+\Lambda}\right)T_{C}
\end{equation}
where $D$ is the size (diameter) of the particle. When the size effects are negligible (that is when $D \gg \Lambda$), from Eq.(5) we obtain $T_{Cb}\simeq T_{C}$. For the highest temperature in our study ($T=350K$), we have $\Lambda \simeq 8nm$ for an estimate of the de Broglie wavelength (assuming a free electron mass for $m$). Therefore, for nanoparticles with an average size of $D=6nm$ (with $D \simeq \Lambda$) the size effects in our ferrofluid can not be neglected leading to a rather strong reduction of the bulk Curie temperature $T_C$. More precisely, given the above estimates, Eq.(5) predicts $T_{Cb}= 0.44T_{C}\simeq 348K$ for the true Curie temperature of a single particle in our sample, in a rather good agreement with the observations.

\section{Conclusions}

In summary, the analysis of the high-temperature behavior of magnetization in $CoFe_2O_4$ based ferrofluid revealed the presence of two phase transitions. The first (major) transition takes place inside single particle. With increasing an applied magnetic field to the saturation level, a second transition starts to develop at lower temperature. Its origin is attributed to manifestation of rather strong proximity type effects between FM nanoparticles separated by non-FM layers.

\section*{Acknowledgements}

The authors thank Dr. L. Fetisov for realization of magnetic measurements. The financial support through the grants of the Romanian Governmental Representative in the Joint Institute for Nuclear Research (JINR) and the UPB-JINR Cooperation Programme of scientific projects for the period 2016-2017 is acknowledged. This work was partially supported by Brazilian agencies CNPq, CAPES and FAPESQ (DCR-PB).

\end{document}